\documentclass[showpacs,preprintnumbers,amsmath,amssymb, nofootinbib]{revtex4}
\newcommand{\be}{\begin{equation}}
\newcommand{\en}{\end{equation}}
\newcommand{\bea}{\begin{eqnarray}}
\newcommand{\ena}{\end{eqnarray}}

\begin{document}

\title{Holographic Cosmological Models on the Braneworld}

\author{Samuel Lepe}
\address{Instituto de F\'{\i}sica, Pontificia
Universidad Cat\'olica de Valpara\'{\i}so, Casilla 4950,
Valpara\'{\i}so, Chile.}
\author{Francisco Pe{\~n}a}

\address{Departamento de Ciencias F\'{\i}sicas, Facultad de
Ingenier\'{\i}a, Ciencias y Administraci\'on,\\ Universidad de la
Frontera, Avda. Francisco Salazar 01145, Casilla 54-D, Temuco,
Chile.}
\author{Joel Saavedra}

\address{Instituto de F\'{\i}sica, Pontificia
Universidad Cat\'olica de Valpara\'{\i}so, Casilla 4950,
Valpara\'{\i}so, Chile.}

\begin{abstract}
In this article we have studied a closed universe which a
holographic energy on the brane whose energy density is described
by $\rho \left( H\right) =3c^{2}H^{2}$ and we obtain an equation
for the Hubble parameter, this equation gave us different physical
behavior depending if $c^2>1$ or $c^2<1$ against of the sign of
the brane tension.

\end{abstract}

%Uncomment for PACS numbers title message
%\pacs{00.00, 20.00, 42.10}
% Keywords required only for MST, PB, PMB, PM, JOA, JOB?
%\vspace{2pc}
%\noindent{\it Keywords}: Article preparation, IOP journals
% Uncomment for Submitted to journal title message
%\submitto{\JPA}
% Comment out if separate title page not required
\maketitle
%\section{\label{sec:level1} Introduction}
\section{Introduction}
From a modern point of view, we have two main ingredient  in order
to deal with the current problems of modern cosmology (late
acceleration of the universe, cosmological constant problem,
etc.). One of this element is the inclusion of extra dimension by
means of the  warped fifth  dimension that were provided a new
scheme to solve the hierarchy problem in particle physics
\cite{Randall:1999ee, Randall:1999vf}. These models allows to
consider the origin of dark energy (the density matter useful to
explain the late acceleration of the universe) from gravitational
theory and changing the standard stress tensor for matter
\cite{Gabadadze:2007dv}
\begin{equation}
G_{\mu \nu }=8\pi G_{N}\left( T_{\mu \nu }^{matter}+T_{\mu \nu
}^{dark-energy}\right) ,  \label{intro1}
\end{equation}
where $G_{\mu \nu }$ is the $4D$ Einstein tensor, $T_{\mu \nu
}^{matter}$ is the stress tensor for matter and $T_{\mu \nu
}^{dark-energy}$ is the stress tensor for dark energy, a new
exotic component that have a negative pressure to match with the
observation. We can modify this description to one more
fundamental, where the matching with the observation is arises
from the gravitational sector,
\begin{equation}
G_{\mu \nu }-\mathcal{K}_{\mu \nu }=8\pi G_{N}T_{\mu \nu
}^{matter}, \label{intro 2}
\end{equation}
where $\mathcal{K}_{\mu \nu }$ denotes a tensor that arise from
the extrinsic curvature, due to the embedding of our brane
universe in the $5D$ bulk. Therefore from the stadium of
cosmologist we can write the modify Friedmann equation as follow,
\begin{equation}
H^{2}=\frac{8\pi G_{N}}{3}f(\rho ),  \label{intro3}
\end{equation}
where the function $f(\rho)$ encoded all geometric modification of
the cosmological equation. In particular the Randall-Sundrum
scenario has gotten a great attention in the last decade
\cite{Randall:1999ee, Randall:1999vf}. From the cosmological point
of view, brane world offers a novel approach to our understanding
of the evolution of the universe. One of the most spectacular
consequence of this scenario is the modification of the Friedmann
equation. In these models, for instance in five dimensions, matter
is confined to a four dimensional brane, while gravity can be
propagated in the bulk,and can feel the extra dimension. From the
perspective of string theory ~\cite{witten}, brane world cosmology
has been a big challenge for modern cosmology . For a review on BW
cosmology see Ref.\cite{rm} and reference therein.

The other main ingredient is the holographic principle, that in
simple words establish that all degrees of freedom of a region of
space in are the same as that of a system of binary degrees of
freedom distributed on the boundary of the region
\cite{Susskind:1994vu}. This point of view, represents an approach
from a consistent theory of quantum gravity (unfortunately not yet
found) in order to clarify the nature of dark energy.  The
Holographic principle says that the number of degrees of freedom
of a physical system should scale with its bounding area rather
than with its volume. Along these lines the literature have been
focused in explain the size of the dark energy density on the
basis of holographic ideas, derived from the suggestions that in
quantum field theory a short distance  a cut-off is related to a
long distance cut-off due to the limit set by the formation of a
black hole \cite{Cohen:1998zx}. From the brane world approach the
holographic principle was implemented in Ref. \cite{Zhang:2007an}
where was studied  the cosmological evolution of the holographic
dark energy in a cyclic universe, generalizing the model of
holographic dark energy proposed Ref. \cite{Li:2004rb}.

The plan of the paper is as follows: In Sec. II we specify the
effective equation on the brane and give some generalities. In
Sec. III we discuss a closed universe with an holographic density
energy on the brane. Finally, we conclude in Sec. IV.

\section{Generalities}
We are going to consider an homogeneous and isotropic 4-brane
described by the FLRW metric, in the case of closed universe where
the gravitational sector of the field equations is described by a
modified Friedmann equation given by (we adopt a unit system where
8$\pi G=1$)
\begin{equation}
3\left(H^{2}+\frac{1}{a^2}\right) =\rho \left( 1\pm \frac{\rho
}{2\lambda }\right) ,\label{hmodif}
\end{equation}
where $H=\dot{a}/a$ is the Hubble parameter with overdot stands
for derivatives with respect to the cosmic time, and signs are
related to positive and negative brane tension ($\lambda$),
respectively.  This modified Friedmann equation can be derived
from Randall Sundrumm model \cite{Randall:1999ee,Randall:1999vf}
after to projected the five dimensional Einstein equations onto a
four dimensional Friedmann brane \cite{Shiromizu:1999wj} (this
kind of modified Friedmann equation can be derived also from
effective loop quantum cosmology \cite{lqc}). If the matter fields
are confined to the brane, they satisfy the standard conservation
equation
\begin{equation}
\dot{\rho}+3H\left( \rho +p\right)  =0, \label{eq.2}
\end{equation}
From Eqs.(\ref{hmodif}) and (\ref{eq.2}) is straightforward to
disguise the Friedmann equation in the standard form,

\begin{eqnarray}
3\left(H^{2}+\frac{1}{a^2}\right) &=&\rho _{eff},  \label{eq.11} \\
\dot{H} &=&-\frac{1}{2}\left[ \rho _{eff}+p_{eff}\right]
+\frac{1}{a^2}, \label{eq.12}
\end{eqnarray}
where the effective quantities are given by
\begin{eqnarray}
\rho _{eff} &=&2\lambda x\left( 1\pm x\right) ,  \label{eq.13} \\
p_{eff} &=&2\lambda x\left[ \omega \left( 1\pm 2x\right) \pm
x\right] , \label{eq.14}
\end{eqnarray}
and $x=\rho /2\lambda $ and $p=\omega \rho $. The effective
equation is reads as follow,
\begin{equation}
1+\omega _{eff}\left( x\right)
=-\frac{2}{3}\frac{\dot{H}-a^{-2}}{H^2+a^{-2}}, \label{effec.17}
\end{equation}
and for the effective barotropic index we get,
\begin{equation}
\omega _{eff}=\frac{p_{eff}}{\rho _{eff}} =\frac{1}{1\pm x}\left[
\left( 1\pm 2x\right) \omega \pm x\right],  \label{br2}
\end{equation}
where the signs are related to positive and negative brane
tension, respectively. Note that the inclusion of the brane in the
theory change the role of the barotropic parameter ($\omega$) by
another effective parameter called $\omega_{eff}$. The equation
(\ref{br2}) can be writes in the following way
\begin{equation}
1+\omega _{eff}\left( x\right) =\left(1\pm\frac{x}{1\pm
x}\right)\left( 1+ \omega\right), \label{eqq.17}
\end{equation}
then is clear that the presence of the brane implies that the
quantity related to observational data is $\omega _{eff}$ being
$\omega$ a bare quantity.

It is interesting to consider the limit that possess
Eq.(\ref{br2}) in the case of negative brane tension. When $x<<1$,
we are recovering the standard Friedmann equation, and the
barotropic index tend to a constant $ \omega _{eff}\left(
x\rightarrow 0\right) \rightarrow \omega $, i.e one recovers the
standard four dimensional general relativity. On the opposite
limit when $x\rightarrow1$ (strong limit) we obtain
\begin{equation}
\omega _{eff}\left( x\rightarrow 1\right) \rightarrow
-\frac{\omega +1}{1-x}, \label{eq.19}
\end{equation}
and it is possible to avoid the singularity if we take  $\omega
=-1$, then $\omega _{eff}\left( x\right) =-1$.
\section{\label{sec:holo}Closed Universe and Holography}
It is well established that in quantum field theory  a short
distances the UV cutoff is related to a long distance IR cutoff
($L$) due to the limit set by forming a black hole
\cite{Cohen:1998zx}. This implies that the total energy of the
system with size $L$ should not exceed the mass of the associated
black hole. This fact traduce that the holographic density must be
satisfied $L^3\,\rho_H \leq L\,M_{p}^2$. Thus the holographic
energy density is chosen as the one that saturating this
inequality and is given by
\begin{equation}
\rho \left( L\right) =\frac{3c^{2}\,M_{p}^2}{8\pi\,L^2},
\label{eq.222}
\end{equation}
where $M_p$ represent the Planck mass and $3c^2$ is a numerical
factor. We consider the Hubble parameter as our IR cutoff
($L=H^{-1}$). Then the holographic energy density is given by
\begin{equation}
\rho \left( H\right) =3c^{2}H^{2},  \label{eq.22}
\end{equation}
where we adopt natural units. If we consider a closed universe,
from Eqs. (\ref{eq.22}) and (\ref{hmodif}) we obtain the following
equation for the Hubble parameter
\begin{equation}
\left( 1-c^{2}\right) H^{2}=\pm \frac{1}{6\lambda }\left(
3c^{2}\right) ^{2}H^{4}-\frac{1}{a^{2}},  \label{eq.23}
\end{equation}
this equation (note that, in the flat case Eq. (\ref{eq.23})
always drives to de Sitter expansion) gives different physical
behavior depending if $c^2>1$ or $c^2<1$ (this occurs because we
have two possible choices of the signs of $c^2$ and $\lambda$).
In the following we focus on this point.

\subsection{Case $c^2<1$ and $\lambda>0$}
If $c^2<1$, the left side of Eq. (\ref{eq.23}) is positive and
this implies that $\lambda>0$ (the option $c^2<1$ and $\lambda<0$
drives to a non-physical behavior, $H^2<0$), and this equation can
be written in the following way
\begin{equation}
\left( 1-c^{2}\right) H^{2}=\frac{1}{6\lambda }\left(
3c^{2}\right) ^{2}H^{4}-\frac{1}{a^{2}}.  \label{eq.24}
\end{equation}
The algebraic solution of  (\ref{eq.24}) is
\begin{equation}
H^{2}\left( a\right) =\alpha \left( 1+\sqrt{1+\beta a^{-2}}\right)
, \label{eq.25}
\end{equation}
where the constants are given by
\begin{equation}
\alpha =\frac{3\lambda \left( 1-c^{2}\right) }{\left( 3c^{2}\right) ^{2}}%
\,\,\,\,,\,\,\,\,\beta =\frac{2}{3}\frac{\left( 3c^{2}\right) ^{2}}{%
\lambda \left( 1-c^{2}\right) ^{2}}\,\,.  \label{eq.26}
\end{equation}
Using  (\ref{eq.25}) is straightforward to get
\begin{equation}
\dot{H}\left( a\right) =-\frac{1}{2}\alpha\, \beta\, a^{-2}\frac{1}{\sqrt{%
1+\beta a^{-2}}}.  \label{eq.27}
\end{equation}
With the definition of redshift $1+z=a_{0}/a$, the acceleration of
the universe gets
\begin{equation}
\dot{H}\left( z\right) +H^{2}\left( z\right) =\alpha \left[ 1+\frac{1+\frac{%
\beta }{2a_{0}^{2}}\left( 1+z\right) ^{2}}{\sqrt{1+\frac{\beta }{a_{0}^{2}}%
\left( 1+z\right) ^{2}}}\right] >0,  \label{eq.28}
\end{equation}
and now in according to Eq. (\ref{br2}) the effective barotropic
index can be written as
\begin{equation}
\omega _{eff}\left( z\right) =\frac{\omega \left[ 1+\frac{\left(
1-c^{2}\right) }{c^{2}}\left( 1+\sqrt{1+\frac{\beta
}{a_{0}^{2}}\left( 1+z\right) ^{2}}\right) \right] +\frac{\left(
1-c^{2}\right) }{2c^{2}}\left(
1+\sqrt{1+\frac{\beta }{a_{0}^{2}}\left( 1+z\right) ^{2}}\right) }{1+\frac{%
\left( 1-c^{2}\right) }{2c^{2}}\left( 1+\sqrt{1+\frac{\beta }{a_{0}^{2}}%
\left( 1+z\right) ^{2}}\right) },  \label{eq.29}
\end{equation}
for early time, its limit limit behavior is
\begin{equation}
1+\omega _{eff}\left( z\rightarrow \infty \right) \rightarrow
2\left(1+\omega \right) ,  \label{eq.30} \\
\end{equation}
and the respective limit for the acceleration
\begin{equation}
\ddot{a}\left( t\rightarrow 0\right) \rightarrow \frac{\alpha \sqrt{\beta }}{%
2}\Rightarrow a\left( t\right) \sim t^{2}\Leftrightarrow \omega
_{eff}\left( z\rightarrow \infty \right) \rightarrow -\frac{2}{3}.
\label{eq.31}
\end{equation}
Therefore, at early times we can say that the effective system
associated with the presence of the brane and the holographic
energy as main ingredients, is described by a membrane gas which
has an effective equation of state given by
$p_{eff}\rightarrow-\frac{2}{3}\rho_{eff}$, and the bare matter
(associated with $\omega$) is doomed to becomes described by a
component whose behavior is near to a cosmological constant
$\omega\sim -1$ behavior, but it does not exactly correspond to
cosmological constant.

And for late time
\begin{equation}
1+\omega _{eff}\left( z\rightarrow -1\right) \rightarrow \left(
2-c^{2}\right) \left( 1+\omega \right),  \label{eq.32}
\end{equation}
and the respective limit for the acceleration
\begin{equation}
\ddot{a}\left( t\rightarrow \infty \right) \rightarrow 2\alpha
a\Rightarrow a\left( t\right) \sim \exp \left( \sqrt{2\alpha
}t\right) \Leftrightarrow \omega _{eff}\left( z\rightarrow
-1\right) \rightarrow -1,  \label{eq.33}
\end{equation}
in this case the effective  and the bare description are exactly
described by a cosmological regimen. It is gives an accelerated
evolution driven by a de Sitter expansion.

Then these scenarios where the presence of the brane is reflected
in the fact that we have positive brane tension and the inclusion
of holographic energy density allows to unify at least two
accelerated phases in the evolution of the universe, described by
a component whose behavior at early time is near cosmological
constant and late time is exactly as a cosmological constant.
Therefore the barotropic bare index $\omega$ has a slow variation
in the neighborhood of cosmological constant domain.

\subsection{$c^2>1$ and $\lambda<0$}
Now if we consider that $c^{2}>1$ implies that $\lambda<0$ (the
option $c^2>1$ and $\lambda>0$ drives to a non-physical behavior,
$H^2<0$), the expression (\ref{eq.24}) is
\begin{equation}
\left( c^{2}-1\right) H^{2}=\frac{1}{6\lambda }\left(
3c^{2}\right) ^{2}H^{4}+\frac{1}{a^{2}},  \label{eq.34}
\end{equation}
and the solutions of this equation are
\begin{equation}
H_{\pm }^{2}(a)=\overline{\alpha }\left( 1\pm \sqrt{1-\overline{\beta }a^{-2}%
}\right) ,  \label{eq.35}
\end{equation}
where the constant are given by
\begin{equation}
\overline{\alpha }=\frac{3\lambda \left( c^{2}-1\right) }{\left(
3c^{2}\right) ^{2}}\,\,\,\,\,and\,\,\,\,\,\overline{\beta }=\frac{2}{3}%
\frac{\left( 3c^{2}\right) ^{2}}{\lambda \left( c^{2}-1\right)
^{2}}. \label{eq.36}
\end{equation}
Such that, from Eq. (\ref{eq.35}) we obtain
\begin{equation}
\dot{H}_{\pm }\left( a\right) =\pm \frac{\overline{\alpha }\overline{\beta }%
}{2}\frac{1}{a^{2}\sqrt{1-\overline{\beta }a^{-2}}},
\label{eq.37}
\end{equation}
and for the acceleration we get
\begin{equation}
\dot{H}_{\pm }\left( a\right) +H_{\pm }^{2}\left( a\right) =\frac{\overline{%
\alpha }}{a^{2}\sqrt{1-\overline{\beta }a^{-2}}}\left[ a^{2}\left( \sqrt{1-%
\overline{\beta }a^{-2}}\pm 1\right) \mp \frac{\overline{\beta
}}{2}\right] . \label{eq.38}
\end{equation}
From the positivity  of the discriminant $1-\overline{\beta
}a^{-2}$ we obtain the following constraint over the redshift
\begin{equation}
\frac{a}{a_{0}}=\left( 1+z\right) ^{-1}>\left( 1+z_{c}\right)
^{-1}, \label{eq.39}
\end{equation}
where the critical redshift value is given by
\begin{equation}
1+z_{c}=\frac{a_{0}}{\sqrt{\overline{\beta }}},  \label{eq.40}
\end{equation}
so the allowed redshift must be satisfy the following cosntraint
\begin{equation}
-1\leq z<z_{c},  \label{eq.41}
\end{equation}
and we note that this constraint exclude the early times limit
gives by $z>z_c$.

Now for the effective barotropic index we are applying a slightly
different approach than the previous section. As the critical
redshift does not allow to take at early times we do not write the
effective index in terms of the bare barotropic index and we just
focus our attention in effective description
\begin{equation}
1+\omega ^{\pm
 }_{eff}\left( z\right) =\mp \frac{1}{3}\Delta ^{\pm }\left(
z\right) ,  \label{eq.42}
\end{equation}
where
\begin{equation}
\Delta ^{\pm }\left( z\right) =\left( \frac{1+z}{1+z_{c}}\right)
^{2}\left[ \sqrt{1-\left( \frac{1+z}{1+z_{c}}\right) ^{2}}\left(
1\pm \sqrt{1-\left( \frac{1+z}{1+z_{c}}\right) ^{2}}\right)
\right] ^{-1},  \label{eq.43}
\end{equation}
such that, $\Delta ^{\pm }\left( z\right) >0$. From
Eqs.(\ref{eq.42}) and (\ref{eq.43}) we obtain two possible schemes
\begin{equation}
\omega^{+}_{eff}\left( z\right) =-1-\frac{1}{3}\Delta ^{+}\left(
z\right) <-1, \label{eq.44}
\end{equation}
this corresponds to a phantom behavior that asymptotically goes to
a genuine de Sitter expansion because $\Delta ^{\pm}\left(
z\rightarrow -1\right)\rightarrow 0$. From the early limit
($z\rightarrow z_c$), we observe a peculiar behavior of our
physical quantities because appears an early singularity, that in
the literature of phantom cosmology is classified as a sudden
singularity, are described by $a$ and $\rho$ finite and $|p|$
infinite \cite{Nojiri:2005sx}.
\begin{equation}
\omega^{-}_{eff}\left( z\right) =-1+\frac{1}{3}\Delta ^{-}\left(
z\right), \label{eq.45}
\end{equation}
that represent a  transitory behavior starting with matter,
passing to a quintaessential transitory phase and an
asymptotically like cosmological constant
$\omega^{-}_{eff}\rightarrow -1$ behaviors. Although
$\omega^{-}_{eff}\rightarrow -1$  the evolution of scalar factor
does not goes as de Sitter expansion ($\ddot{a} \sim
a^{-1},\,z\rightarrow-1$). Also this branch have a sudden early
singularity but that does not correspond to a phantom singularity.

Summarize in a scenario where the brane tension is negative and
the holographic parameter satisfies $c^2>1$, we obtain two
branches that driven to different cosmological stages. First at
all, we obtain a critical value over the redshift ($z_c$), that
constraint the physical region of our branches, where one of them
represent an  early phantom evolution goes asymptotically to a
genuine de Sitter and the other one correspond to a early matter
stage, then a transitory quintaessential phase and ending in an
asymptotically like cosmological constant behavior. Also we want
to note that both branches beginning with a early sudden
singularity, where one of them correspond to a phantom singularity
and the other does not generate a phantom. The difference lies in
the early limit of the effective barotropic index  because $\omega
^{\pm}_{eff}\rightarrow \mp \infty$.

\section{Discussion and Outlook}
In this article we have studied a closed universe which a
holographic energy on the brane whose energy density is described
by $\rho \left( H\right) =3c^{2}H^{2}$ and we obtain an equation
for the Hubble parameter, given by
\begin{equation}
\left( 1-c^{2}\right) H^{2}=\pm \frac{1}{6\lambda }\left(
3c^{2}\right) ^{2}H^{4}-\frac{1}{a^{2}},  \label{eq.23}
\end{equation}
this equation gave different physical behavior depending if
$c^2>1$ or $c^2<1$ against of the sign of the brane tension. First
we focus on the case $c^2<1$ and $\lambda>0$, where the presence
of the brane is reflected in the fact that we have positive brane
tension and the inclusion of holographic energy density allow to
unify at least two accelerated phases in the evolution of the
universe, described by a component whose behavior at early time is
near cosmological constant and late time is a cosmological
constant. Therefore the barotropic bare index $\omega$ has a slow
variation in the neighborhood of cosmological constant domain. In
the other case $c^2>1$ and $\lambda<0$ where the brane tension is
negative and the holographic parameter that satisfies $c^2>1$, we
obtain two branches that driven to different cosmological stages.
First at all, we obtain a critical value over the redshift
($z_c$), that constraint the physical region of our branches,
where one of them represent an  early phantom evolution goes
asymptotically to a genuine de Sitter and the other one correspond
to a early matter stage, then a transitory quintaessential phase
and ending in an asymptotically like cosmological constant
behavior. Also we want to note that both branches beginning with a
early sudden singularity, where one of them correspond to a
phantom singularity and the other does not generate a phantom. The
difference lies in the early limit of the effective barotropic
index  because $\omega ^{\pm}_{eff}\rightarrow \mp \infty$.
Therefore the inclusion of the brane world and holographic energy
on the brane allow to obtain physical results when  we choose the
IR cutoff as the size of our universe ($L=1/H$). We hope to
discuss in the near future the inclusion of the other two
possibilities for $L$ \cite{Li:2004rb}.

\section*{Acknowledgments} This work was supported by COMISION
NACIONAL DE CIENCIAS Y TECNOLOGIA through FONDECYT \ Grant
11060515 (JS). This work was also partially supported by PUCV
Grants No. 123.701/2008 (SL), No. 123.789/2007 (JS) and by
Direcci\'on de Estudios Avanzados PUCV. Also was supported from
DIUFRO DI08-0041 of Direcci\'on de Investigaci\'on y Desarrollo
Universidad de la Frontera (FP) . The authors SL and JS wish to
thank Departamento de F\'{\i}sica de la Universidad de La Frontera
for its kind hospitality.

\end{document}